\RequirePackage{fix-cm}
\documentclass[smallextended]{svjour3}       
\smartqed  
\usepackage{graphicx}
\usepackage[utf8]{inputenc}
\usepackage{amsmath,amssymb}
\usepackage{tikz}
\usepackage[normalem]{ulem}
\usepackage{indentfirst}
\usepackage[T1]{fontenc}
\usepackage{graphics}
\usepackage{xcolor}
\usepackage{epigraph, varwidth}
\usepackage{pdfpages}
\usepackage{setspace}
\usepackage[font=small,labelfont=bf]{caption}
\usepackage{float}
\usepackage{pgfplots}
\pgfplotsset{compat=1.18} 
\usepackage{epstopdf}
\usepackage{subfig}
\usepackage{afterpage}
\usepackage{array}
\usepackage{emptypage}
\usepackage[retainorgcmds]{IEEEtrantools}
\usepackage[acronym,nonumberlist,sort=use]{glossaries}
\usepackage{enumitem}
\usepackage[colorlinks=true, allcolors=black, citecolor=blue, urlcolor=blue]{hyperref}
\usepackage[super]{nth}
\usepackage{natbib}
\setcitestyle{aysep={}} 
\newcommand{\be}{\begin{equation}}
\newcommand{\ee}{\end{equation}}
\newcommand{\bal}{\begin{aligned}}
\newcommand{\eal}{\end{aligned}}
\newcommand{\beq}{\begin{eqnarray}}
\newcommand{\eeq}{\end{eqnarray}}
\newcommand{\ba}{\begin{array}}
\newcommand{\ea}{\end{array}}

\newcommand{\bi}{\begin{itemize}}
\newcommand{\ei}{\end{itemize}}

\AtBeginDocument{%
  \paperwidth=\dimexpr
    1in + \oddsidemargin
    + \textwidth
    + 1in + \oddsidemargin
  \relax
  \paperheight=\dimexpr
    1in + \topmargin
    + \headheight + \headsep
    + \textheight
    + 1in + \topmargin
  \relax
  \usepackage[pass]{geometry}\relax
}

\begin{document}

\title{
Multistability and Gibbs entropy in the planar dissipative spin-orbit problem
}
%



\author{Vitor M. de Oliveira
}



\institute{\at Instituto de Matemática e Estatística, Universidade de São Paulo, 05508-090 São Paulo/SP, Brazil, and \at CFisUC, Departamento de Física, Universidade de Coimbra, 3004-516 Coimbra, Portugal \\ 
\email{vitormo@ime.usp.br | vitor.m.oliveira@uc.pt}
}

\date{Received: date / Accepted: date}

\maketitle

\begin{abstract}
In this work, we numerically investigate and visually illustrate the dynamical properties of the dissipative spin-orbit problem such as the co-existence of multiple periodic and quasi-periodic attractors, and the complexity of the corresponding basins of attraction. Our model is composed by a triaxial satellite (planet) orbiting a planet (star) in a fixed Keplerian orbit with zero obliquity. A dissipative tidal torque that is proportional to its rotational angular velocity is assumed to be acting on the satellite. We use Hyperion as a toy model to characterize the methodology used, since this system has a very rich conservative dynamical scenario, and we later apply our methodology to the Moon and Mercury. Our results show that the basins of attraction may possess an intricate structure in all cases which changes with the orbital eccentricity, and that the Gibbs entropy is a good measure on how dominant one basin is over the others in the phase space.

\end{abstract}

\section{Introduction}
\label{sec:intro}

Many natural satellites in the Solar System, including the Moon, are observed in an orbital period similar to their rotational period. These bodies probably rotated much fast upon their formation and then went through a despinning process due to tidal interactions with their massive companions \citep{Macdonald1964, Kokubo2007}, until eventually ending up in a synchronous rotation. Since the seminal work by \cite{Darwin1879I}, many authors have studied the rotational dynamics of orbiting bodies under the effect of tidal forces, such as \cite{Efroimsky2012, Ferraz2013, Correia2014, Ragazzo2017, Correia2019}. Such phenomenon mainly affects the motion of orbiting bodies that are bigger or closer to the central body \citep{Murray1999}, but it can also impact the evolution of small bodies \citep{Goldreich2009}.

The closest planet to the Sun, Mercury, is currently moving in a 3/2 spin-orbit resonance, i.e., it takes the same amount of time to rotate around its spin axis thrice as it takes to revolve around the Sun twice. This discover by \cite{Pettengill1965} showed that the system can be locked into other configurations besides the synchronous one, which has lead to many works regarding the final state of tidally evolved systems. In particular, \cite{Goldreich1966} derived an analytical probability for a planet or satellite to be captured into a spin-orbit resonance based on arguments about energy dissipation, while \cite{Henrard1985} reinterpreted the probability of capture in terms of Adiabatic Invariant Theory. Later, \cite{Correia2004} showed that the chaotic evolution of Mercury's orbit could have driven its orbital eccentricity high enough during the planet's history so that the probability of it ending up in its current state is actually higher than previously thought.

From a Dynamical Systems perspective, we can interpret the probability of being captured by a resonance as the probability of a random trajectory converging to a given final state in the phase space. In a system with multistability, i.e., many co-existing attractors, such concept can be related to the size of the basin of attraction associated with each attractor (see \citeauthor{Celletti2008} \citeyear{Celletti2008}). Here, we are interested in visually illustrating and also quantifying the complexity of the basins of attraction in the spin-orbit problem. To carry on with this investigation, we choose a low-dimensional model of the problem which assumes that the orbiting body follows a Keplerian orbit with fixed parameters and rotates perpendicularly to the orbit plane. We then determine the basins of attraction of the main resonances, and we measure their respective sizes and entropy. Our results highlight the rich dynamical scenarios that may emerge from such a system. We depict the intricate structure formed by the basins of attraction in the spin-orbit problem, and we evaluate how dominating one of the attractors is over the others via the Gibbs entropy, which are the main contributions of this work.

This paper is organized as follows. In Sec. \ref{sec:system} we present the equations of motion for our physical model. Later, in Sec.~\ref{sec:entropy} we briefly expose the main ideas regarding the Gibbs entropy in our context. The dynamical analyses for Hyperion, the Moon and Mercury are presented in Secs.~\ref{sec:hyperion}~and~\ref{sec:moon_and_mercury}. Finally, in Sec. \ref{sec:conclusions} we give some general remarks and state our conclusions.

\section{Physical model}
\label{sec:system}

Our physical model consists in an orbiting body that revolves around a central body with the following considerations: the orbit is given by a fixed Keplerian ellipse with eccentricity $e$, semi-major axis $a$ and instantaneous radius $r$; the central body is a point particle whilst the orbiting body is a triaxial almost-rigid rotating body with principal moments of inertia $I_1<I_2<I_3$; the orbiting body's spin-axis is parallel to $I_3$ and perpendicular to the orbit plane; and the only forces acting on the orbiting body are due to the gravitational field generated by the central body.

We also consider that the central body rises a tide on the orbiting body. The time lag $\tau$ between the distortion of the body and the tide-raising potential is assumed to be constant \citep{Singer1968, Mignard1979}, which leads to a phase lag, and consequently a dissipation function, that is linearly dependent on the relative angular velocity $\dot{\psi}$ \citep{Peale2005, Efroimsky2007}.

\begin{figure*}[h]
\centering
\includegraphics[scale=0.4]{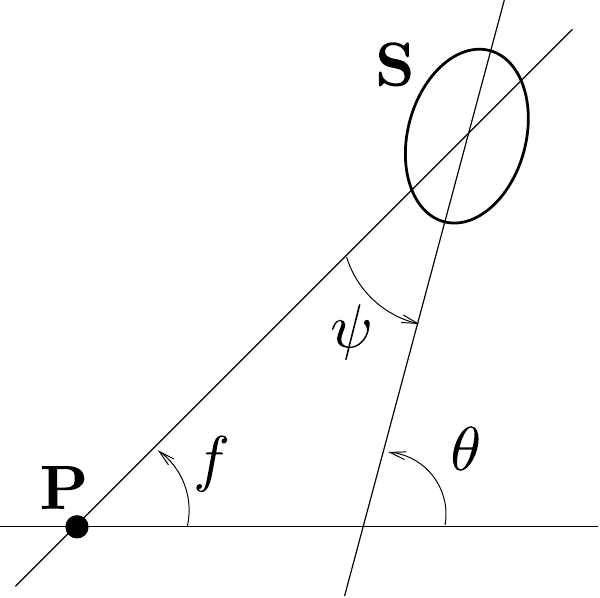}
\caption{Schematic of the system. The satellite ${\bf S}$ orbits the planet ${\bf P}$ following a Keplerian orbit. $f$ stands for the true anomaly and $\theta$ for the rotation angle.}
\label{fig:schematic}
\end{figure*}

The position of the orbiting body along its trajectory may be determined by the true anomaly $f$, and its rotation may be evaluated by the angle $\theta$ between the axis along $I_1$ (largest physical axis) and a coordinate axis fixed in inertial space, which we take to intersect both the central body and the point of periapsis. We illustrate the system in Fig.~\ref{fig:schematic}.

With the aforementioned considerations, our equation of motion is

\begin{equation}
  I_3 \ddot \theta = -\frac{3}{2}(I_2-I_1)\frac{Gm_P}{r^3}\sin 2\psi-3 k_2 \frac{Gm_P^2 R^5}{r^6}\tau\dot \psi,
  \label{eq:motion_full}
\end{equation}

\noindent where $G$ is the gravitational constant, $m_P$ is the planet's mass and \mbox{$\psi=\theta-f$}. The first term on the right-hand side of Eq.~\eqref{eq:motion_full} is the gravitational torque, and the second term is the dissipative tidal torque, where $k_2$ is the Love number and $R$ is the satellite's radius.

By rearranging the terms in Eq.~\eqref{eq:motion_full} and denoting the equatorial oblateness by $\gamma=3(I_2-I_1)/2I_3$, our equation of motion can be rewritten as

\begin{equation}
  \ddot \theta = -\gamma\frac{Gm_P}{r^3}\sin 2(\theta-f)-K\big(L \dot \theta-N\big),
  \label{eq:motion}
\end{equation}
\noindent with
\begin{equation}\nonumber
    L = \dfrac{a^6}{r^6} \qquad \text{and} \qquad N = \dfrac{a^8n\sqrt{1-e^2}}{r^8},
\end{equation}
\noindent where $n$ is the satellite's mean motion.

We are interested in spin-orbit resonances, solutions where the number of revolutions around the central body and the number of rotations around its spin-axis are commensurable. Following the definition in \cite{Celletti2008}, if $T$ is the satellite's orbital period, a \emph{spin-orbit resonance} (SOR) of type $p/q$ is a solution for the system such that
\begin{equation}\nonumber
 \theta(t+ T q)=\theta(t)+ 2\pi p \qquad \text{for all} \qquad t\in\mathbb{R},
\end{equation}
for two relatively prime integers $p$, $q$  with $q > 0$.

The spin-orbit resonances are, by definition, periodic orbits. For $K=0$, Eq.~\eqref{eq:motion} describes an one-and-a-half degree of freedom near integrable conservative system, with both $\gamma$ and $e$ as perturbation parameters. Therefore, the system can possess a mixed phase space, with the coexistence of regular and chaotic motion \citep{Lichtenberg1992}. For $K>0$, some spin-orbit resonances may become periodic attractors of the system, being the final state of a set of trajectories with positive measure. We call the set of initial conditions that converge in time to a given attractor the \emph{basin of attraction} of said solution \citep{Alligood2012}. For certain parameters, Eq.~\eqref{eq:motion} also has solutions where $p/q$ is irrational. Such orbits form KAM curves in the conservative case and may become quasi-periodic attractors in the dissipative case \citep{Celletti2009}. 

In order to help us analyze the spin-orbit problem, we define a \emph{stroboscopic map} $\bf{M}$ which acts on the system's phase space as $(\theta(t),\dot{\theta}(t))\to(\theta(t+T),\dot{\theta}(t+T))$. With this definition, we reduce the three-dimensional continuous system to a two-dimensional autonomous discrete system. Note that the spin-orbit resonances are represented on $\bf{M}$ by periodic orbits with period $q$, while the quasi-periodic solutions are represented by one-dimensional curves.

The model presented in this Section is well established and was used to study spin-orbit coupling in many works. See, for example, \cite{Wisdom1984, Correia2004, Celletti2008, Celletti2014}. The fact that it is a low-dimensional model gives us access to numerical tools that are capable of visually illustrating the dynamical properties of orbits in the phase space.

\section{Entropy}
\label{sec:entropy}

The basin of attraction is the portion of the phase space where initial conditions lead to orbits that converge to the attractor. Hence the size of the basin of attraction can be interpreted as the probability of a random initial condition to belong in said basin. In multistable systems, multiple basins coexist and their basins compete for room in the phase space. In this situation, the individual basin sizes give the probability of a random initial condition to converge to a certain final state. Although very relevant, this information alone does not reflect how dynamically complex the phase space is. More specifically, it does not show how the phase space is divided among the basins.

In order to infer some information on that matter, we use the Gibbs entropy $S$, which is defined as \citep{Gibbs2010}
\begin{equation}
    S=\sum_{i=1}^{N_A}p_i\ln\left(\dfrac{1}{p_i}\right),
\end{equation}
\noindent where $N_A$ is the number of attractors and $p_i$ is the probability of a given orbit to belong to the basin of attraction of the $i$-th attractor. 

The entropy $S$ condenses in one number how heterogeneous the basin sizes are. It is maximum when all basins have the same size, i.e., $p_i=1/N_A$ for every attractor. In this situation, we obtain
\begin{equation*}
S_{max}=\ln N_A.
\end{equation*}
\noindent Hence, the entropy reflects how dominant an attractor is in the phase space in terms of its basin size, and it also takes into consideration the total number of attractors,  expressed by its maximum value, which depends on the system parameters.

It is worth mentioning that an entropy called \emph{basin entropy} was defined in \citep{Daza2016} as a tool to assist in the classification of boundaries between the basins of attraction, which are related to the uncertainty in dynamical systems. This quantity was based on the Gibbs entropy and it was later shown that it was also related to dynamical and geometrical aspects such as uncertainty exponents and lacunarity \citep{Daza2022}.

As we are going to depict in the next sessions, the structure of the basin boundaries in the dissipative spin-orbit problem can be mostly perceived visually as riddled basins (see, for example, \citeauthor{Alexander1992} \citeyear{Alexander1992}), where the unpredictability is maximum and the measure of uncertainty exponents do not give much information \citep{Daza2022}. Hence, the measure of the system's complexity here can be taken directly from the Gibbs entropy.

\begin{figure*}[htbp]
\centering
  \includegraphics[scale=0.13]{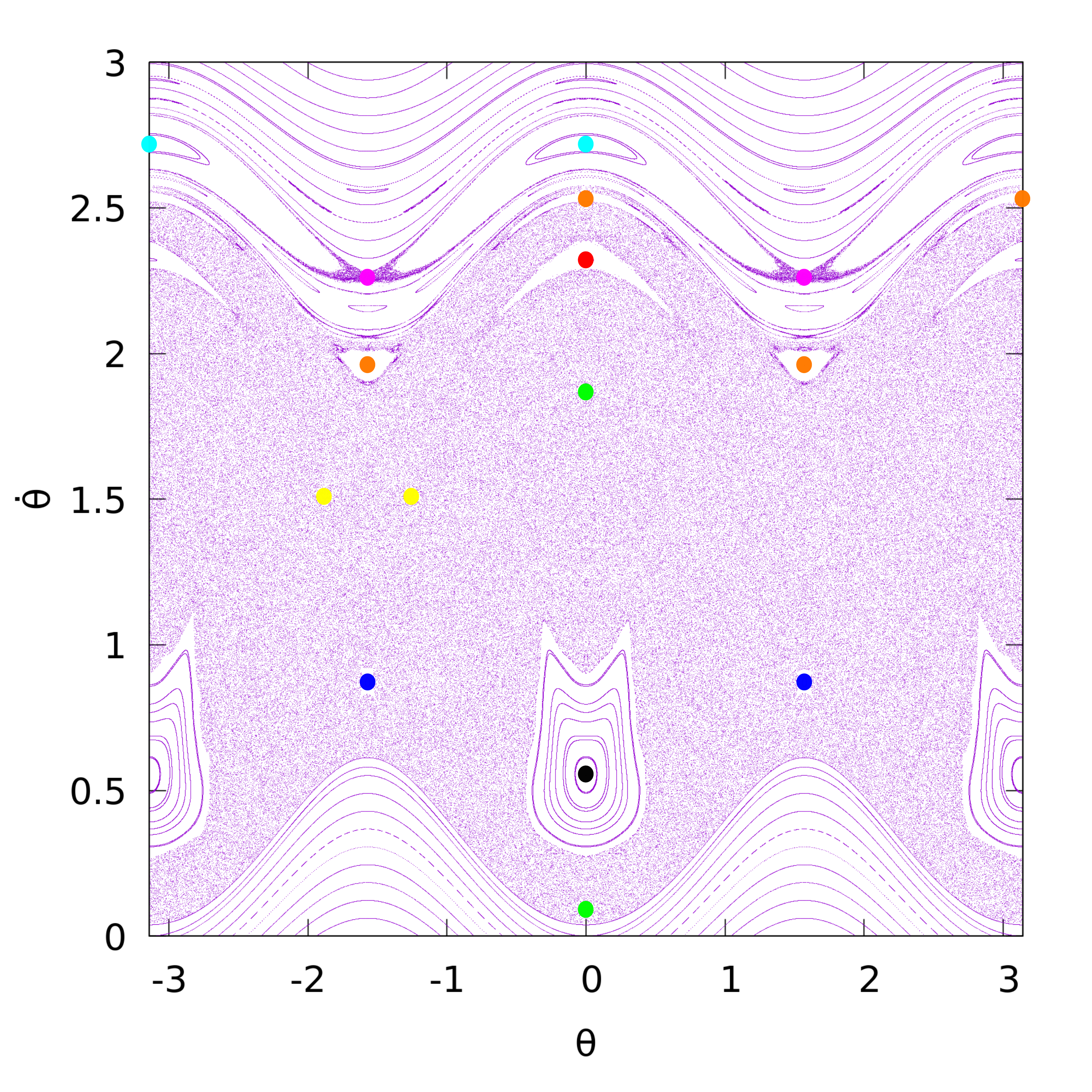}
\caption{Hyperion's phase space with main SORs. 1/1 stable (black), 2/1 stable (red), 1/2 stable (dark blue), 2/2 unstable (green), 3/2 unstable (yellow), 5/2 stable (light blue), 5/2 unstable (pink), and 9/4 stable (orange).}
\label{fig:phase_space_hyperion}
\end{figure*}

\section{Hyperion}
\label{sec:hyperion}

Hyperion is a moon of Saturn that is currently believed to be rotating chaotically \citep{Wisdom1984, Wisdom1987}. This natural satellite has a highly aspherical shape, being nearly twice as long as it is across, with physical parameters given by $e_{hyp}=0.1$ and ${\gamma}_{hyp}=0.396$ \citep{Wisdom1987Urey}. In this section, we adopt these parameters in order to study the spin-orbit problem. 

For all our simulations, we start the orbital motion of the satellite at \emph{periapsis}, i.e., $f(t=0)=0$ and $r(t=0)=a$, and we calculate $r(t)$ via Newton's equation. We also take $a=1$ and $T=2\pi$, which leads to $n=1$ and $Gm_p=1$. Numerical integration is carried out using an explicit embedded Runge-Kutta Prince-Dormand 8(9) scheme with adaptive step-size control \citep{Galassi2001}.

\subsection{Conservative scenario}

Let us first consider the case where Hyperion is treated as a rigid body and hence there is no tidal dissipation in the satellite's rotation. In this case, we have $K=0$, and the gravitational torque is the only term present in Eq.~\eqref{eq:motion}. Figure~\ref{fig:phase_space_hyperion} shows the system's phase space $\theta \times \dot{\theta}$ calculated on the stroboscopic map $M$. We can observe a typical near-integrable Hamiltonian system scenario, with regions of regular motion coexisting with chaos.

\begin{figure*}[htbp]
\centering
  \includegraphics[scale=0.75]{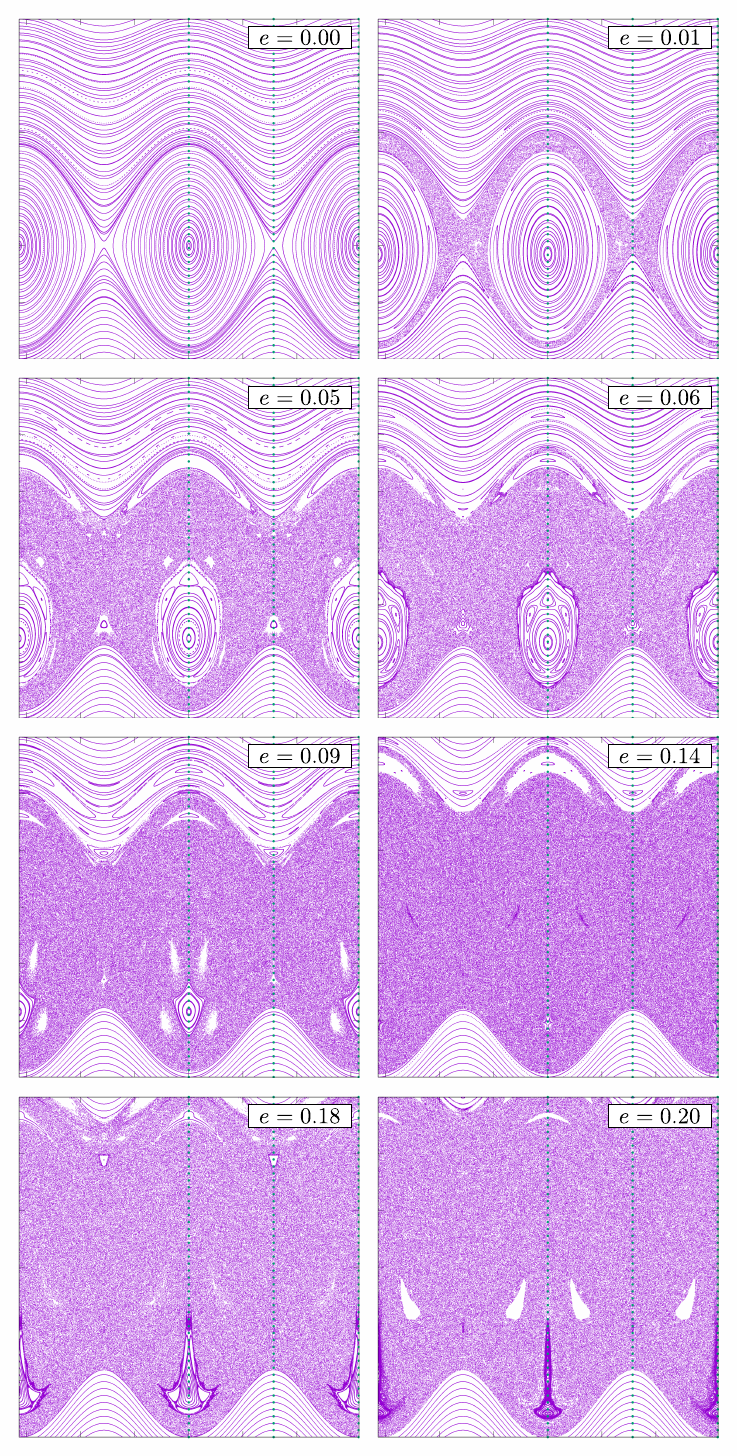}
\caption{Phase space for $\gamma={\gamma}_{hyp}$, $K=0$, and different values of $e$. The initial conditions chosen for these figures are marked by green dots.}
\label{fig:phase_space_hyperion_e}
\end{figure*}

The location of a few low-order spin-orbit resonances are also marked in Fig.~\ref{fig:phase_space_hyperion} by coloured points. We determine these periodic orbits to a high accuracy ($\sim10^{-10}$) using a routine to select good candidates along with an iterative minimization procedure (see, for example, the Appendix of \citeauthor{Raphaldini2020} \citeyear{Raphaldini2020}). For a SOR of type $p/q$, the value of $q$ reflects the period of the orbit on $\bf{M}$, while $p$ is the number of rotations measured outside the surface of section after $q$ intersections.

Since in reality the physical parameters are not constant in time, it is important for us to understand how the system changes as we vary these. In Fig.~\ref{fig:phase_space_hyperion_e}, we present the phase space for different values of the orbital eccentricity $e$ inside a given range. What we observe is the system going from a pendulum-like regular motion to a mixed phase space as the value is increased. During this process, some SORs bifurcate, changing stability, and regions of regular motion vary in size. For even larger values, we have the selected region of the phase space almost fully covered by chaotic orbits.

\subsection{Time series with tidal dissipation}

\begin{figure*}[h]
\centering
  \includegraphics[scale=0.2]{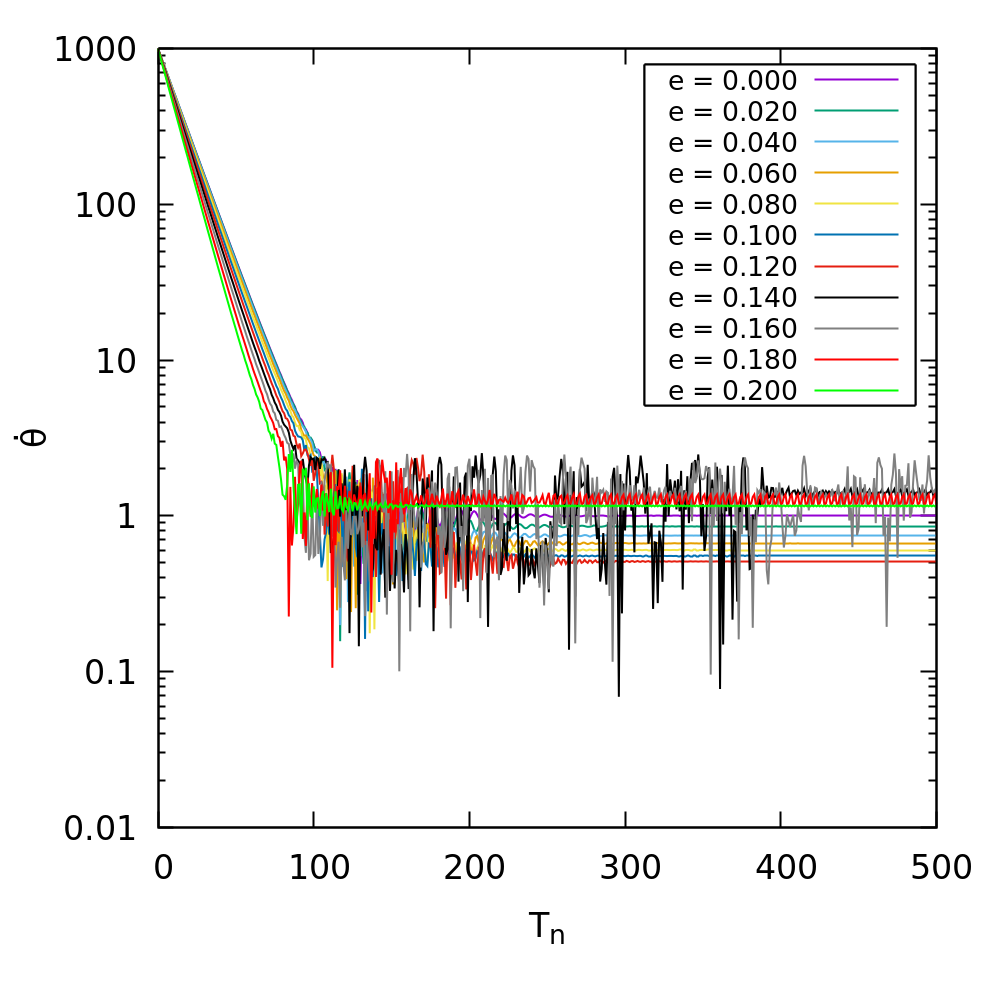}
\caption{Discrete-time series for the angular velocity $\dot{\theta}$ of a trajectory on $\bf{M}$ beginning at $\theta(0)=0.0$ and $\dot{\theta}(0)=1000.0$ rad/s, for $K=10^{-2}$, $\gamma=\gamma_{hyp}$, and different values of $e$.}
\label{fig:time_series}
\end{figure*}

When we consider Hyperion as an almost rigid body, tidal dissipation comes into effect and some of the spin-orbit resonances become attractors. However, as we showed in Fig.~\ref{fig:phase_space_hyperion_e}, the system's dynamical scenario vastly changes as we vary the eccentricity. In Fig.~\ref{fig:time_series}, the evolution of the angular velocity $\dot{\theta}$ is shown as a function of the number of orbital periods $T_n$ (the unity of time in $\bf{M}$), for an initial condition given by $\theta(0)=0.0$ and $\dot{\theta}(0)=1000.0$ rad/s (i.e., Hyperion initially facing Saturn and rotating fast), and for different values of $e$.

We choose a dissipation constant of $K=10^{-2}$, a value which is high enough to allows us to observe the effects that tidal dissipation has on the system in a reasonable amount of simulation time, and low enough so that the interesting dynamical phenomena are not totally suppressed by the dissipation.

Two conclusions can be taken from Fig.~\ref{fig:time_series}. First, the discrete-time series is composed by three parts: an exponential decay, an erratic motion, and an asymptotic state corresponding to a resonance. Second, varying the value of the eccentricity affects the time series mostly in two manners: it changes the slope corresponding to the initial exponential decay, and it possibly modifies the trajectory's final state, i.e., to which SOR the trajectory eventually converges.

When dissipation is considered, the chaotic nature of the system translates as chaotic transients in the phase space. This can be seen in the erratic portion of the time series in Fig.~\ref{fig:time_series}. Hence, the system still presents sensitive dependence on initial conditions, i.e., initial conditions that are close together can lead to very different trajectories, which might converge to different SORs (see, for example, \citeauthor{Ott2002} \citeyear{Ott2002}). Chaotic orbits can also mimic dynamical aspects of unstable periodic orbits inside the chaotic sea for a limited amount of time. With this, a trajectory on the phase space can become temporarily trapped around attractors during its chaotic transient.

Both dynamical aspects mentioned before, namely, sensitive dependence on initial conditions and temporary entrapment, are related to invariant manifolds of unstable periodic orbits (see, for example, \citeauthor{Oliveira2020} \citeyear{Oliveira2020}). These geometrical structures permeate the phase space, bending and crossing each other, and rendering the system dynamics very complex. They also play an important role on the spin-resonance capture, since they form the boundaries of the basins of attraction.

\begin{figure*}[t]
\centering
  \includegraphics[scale=0.9]{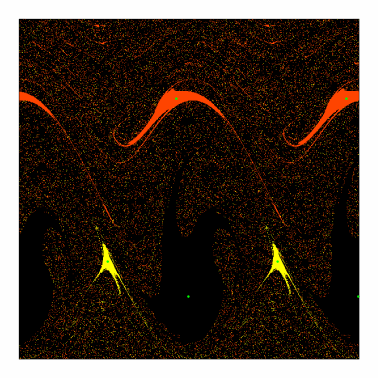}
\caption{Basins of attraction for Hyperion with dissipation constant given by $K=10^{-2}$. The color code corresponds to the following SORs: 1/1 (black), 2/1 (orange), and 1/2 (yellow). The location for the SORs is marked with little green dots.}
\label{fig:boa_hyperion}
\end{figure*}

\subsection{Basins of attraction}

In order to visually illustrate the basins of attraction, we selected a square grid of $600\times600$ initial conditions, evolved the equations of motion for each one and determined their final state. The results for Hyperion are shown in Fig.~\ref{fig:boa_hyperion}. For this system, there are three SORs, namely, 1/1, 1/2, and 2/1, and it is possible to observe two aspects: the basin of the synchronous resonance dominates the phase space, and the boundaries between the basins are complex. The basin sizes, which are based on the number of points on the grid that compose each of them, are presented in Table~\ref{tab:basin_sizes_hyperion}.

\renewcommand{\arraystretch}{1.2}
\begin{table}[htbp]
\caption{Basin sizes and entropy for $e=e_{hyp}$, $\gamma=\gamma_{hyp}$, and $K=10^{-2}$.}
\label{tab:basin_sizes_hyperion}  
\centering
\begin{tabular}{cccc}
\hline\noalign{\smallskip}
\multicolumn{3}{c}{Basin size} & Entropy \\
\hline\noalign{\smallskip}
1/1 & 2/1 & 1/2  &  \\
\noalign{\smallskip}\hline\noalign{\smallskip}
89.94\% & 8.00\% & 2.06\% & 0.344 \\
\noalign{\smallskip}\hline
\end{tabular}
\end{table}

\begin{figure*}[htbp]
\centering
  \includegraphics[scale=0.75]{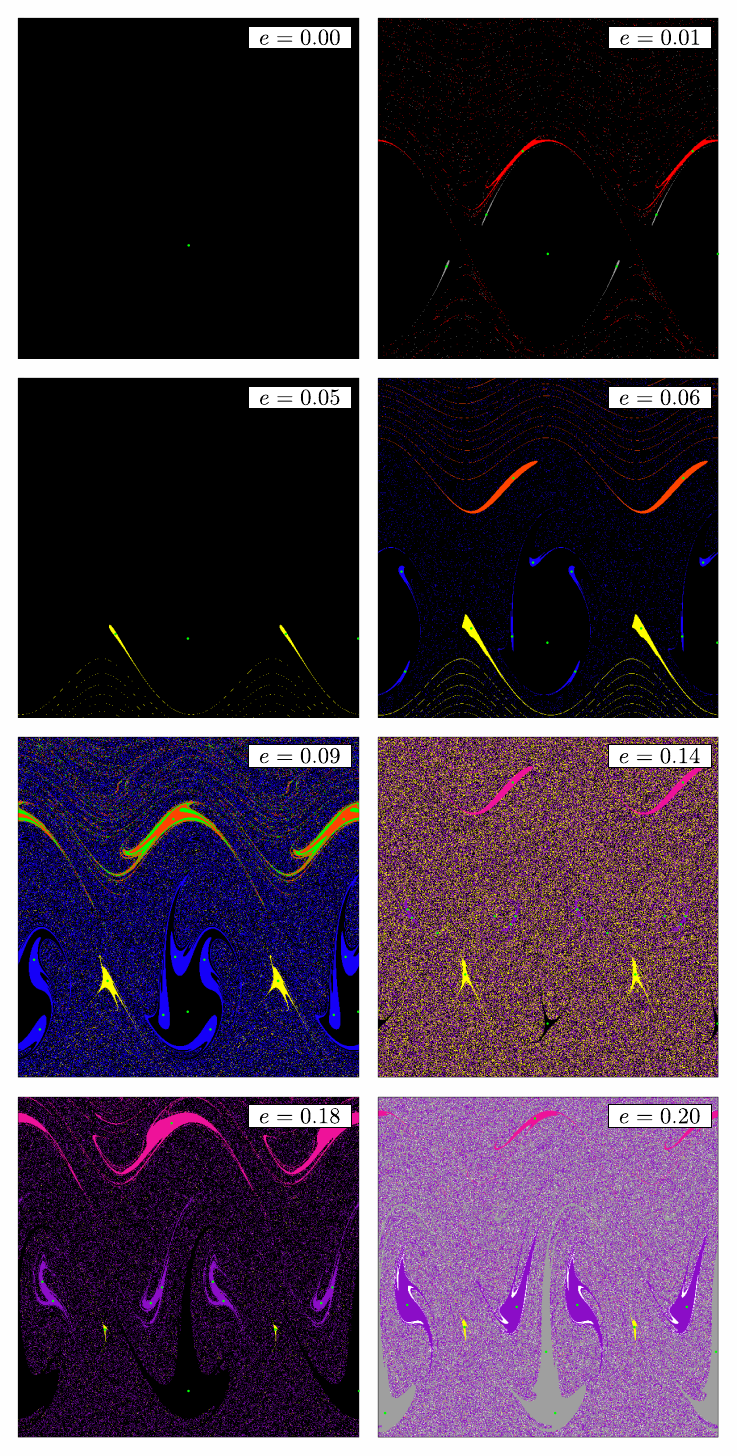}
\caption{Basins of attraction for $K=10^{-2}$, ${\gamma}=\gamma_{hyp}$, and different values of $e$. The color code corresponds to the following SORs: 1/1 (black), 2/1 (orange), 1/2 (yellow), 2/2 (gray), 3/2 (red), 4/2 (dark-green), 5/2 (pink), 4/4 (blue), 5/4 (violet), and 8/4 (green). The basin of higher-order resonances is white, and the location for the SORs is marked with little green dots.}
\label{fig:boa_hyperion_e}
\end{figure*}

In Fig.~\ref{fig:boa_hyperion_e}, we maintain the values for $\gamma$ and $K$ used in Fig.~\ref{fig:boa_hyperion}, and we calculate the basin of attraction of SORs with $p\leq9$ and $q\leq4$ for the same values of $e$ as in Fig.~\ref{fig:phase_space_hyperion_e}. For $e=0$, the $1/1$ SOR is the only resonance present and hence its basin of attraction (in black) fills the phase space. As the value for the orbital eccentricity is increased, other attractors appear along with their basins (multistability). The basins of attraction are notable intertwined with each other in the phase space, highlighting the complex dynamical scenario of the system. For $e=0.2$, the synchronous resonance has bifurcated (period-doubling), giving origin to a $2/2$ SOR (basin in blue). Note that the $2/2$ SOR is different from the $1/1$ SOR, since this solution only closes after the body fully rotates around itself two times, which takes two orbital periods. 

Here we are treating different attractors that correspond to a similar SOR as effectively the same attractor, since we can assume that they lead to the same physical outcome. For example, if the orbiting body is in the synchronous solution, it only shows the same side to the central body. However, there is another synchronous solution, where the orbiting body would always show its other side to the central body. If we were not to consider that these solutions were the same, the complexity of the system would increase since there would be more final states and, consequently, more basins of attraction.

\begin{figure*}[htbp]
\centering
  \includegraphics[scale=0.2]{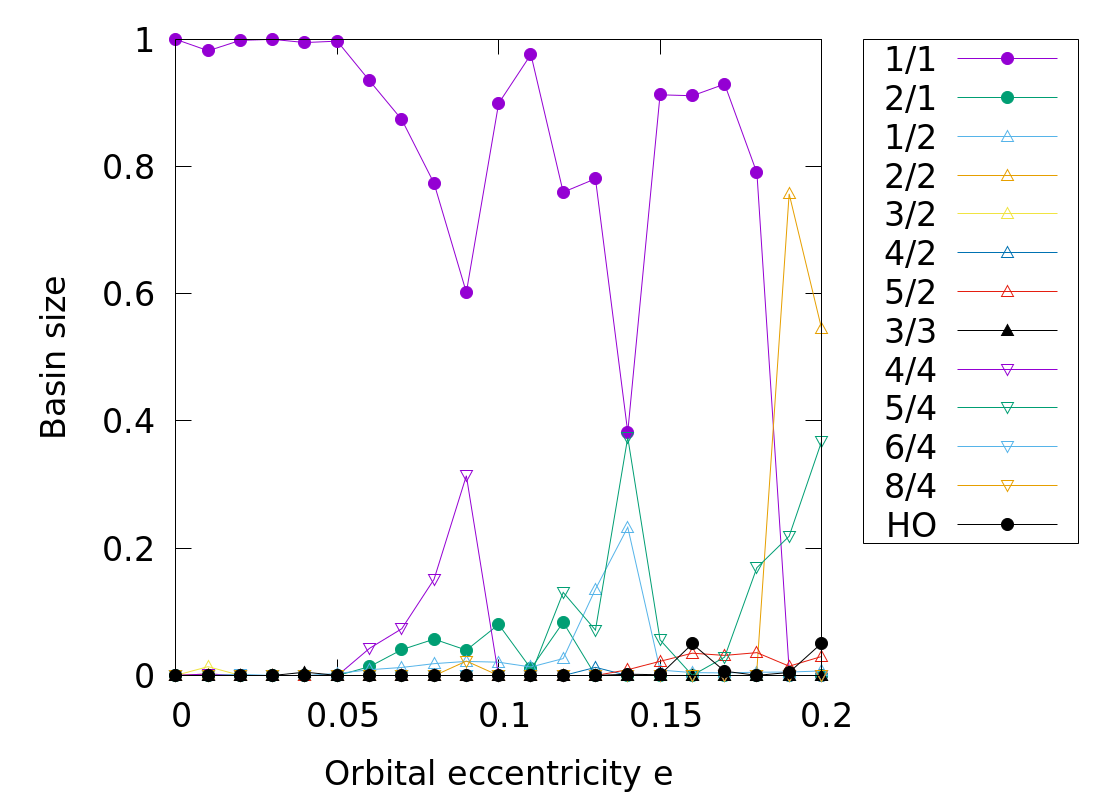}
\caption{Size of the basins of attraction for ${\gamma}=\gamma_{hyp}$ and $K=10^{-2}$. HO stands for higher-order SORs ($p>9$ or $q>4$).}
\label{fig:BOA_size}
\end{figure*}

In Fig.~\ref{fig:BOA_size}, we plot the size of the basins of attraction as a function of the eccentricity $e$. Here, we consider SORs with $p>9$ or $q>4$, higher-order (HO) resonances, as one single final state. As observed in Fig.~\ref{fig:boa_hyperion_e}, the basin size for the synchronous resonance dominates for lower values of eccentricity ($e\leq0.05$). However, it competes for space with other basins as the eccentricity increases, especially for $e=0.09$ and $e=0.14$, where the probability of a random trajectory to converge to the $1/1$ spin-orbit resonance falls to approximately $0.6$ and $0.4$, respectively. For these parameters, the $4/4$ SOR ($e=0.09$), and the $5/4$ and $1/2$ SORs ($e=0.14$) also become probable outcomes.

\begin{figure*}[htbp]
\centering
\includegraphics[scale=0.2]{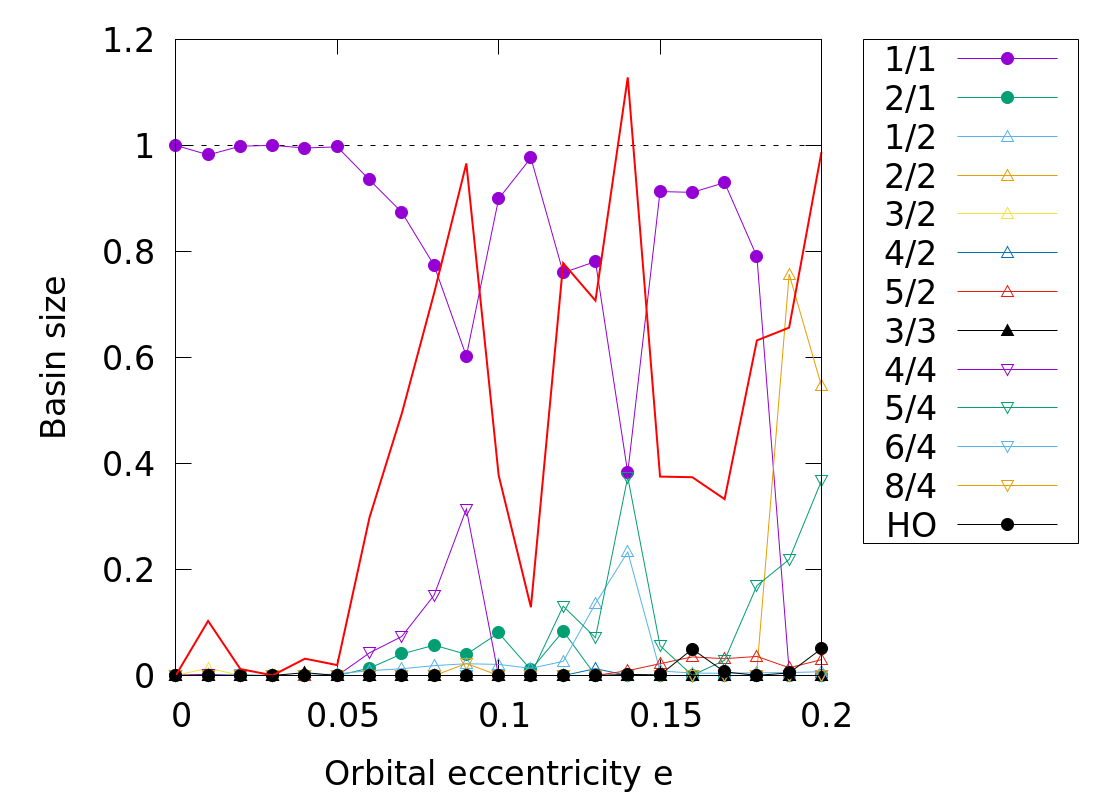}\label{subfig:size_entropy_all}
\caption{Entropy (red curve) as a function of the orbital eccentricity alongside the basin sizes, for ${\gamma}=\gamma_{hyp}$ and $K=10^{-2}$.}
\label{fig:size_entropy}
\end{figure*}

The entropy for the parameters used in Fig.~\ref{fig:BOA_size} are presented in Fig.~\ref{fig:size_entropy}. The most striking result we observe is that the system's entropy is highly correlated to the basin size of the synchronous resonance. As the orbital eccentricity changes, the entropy varies and such variation inversely follows the tendency of the $1/1$ SOR basin size, even when its value goes below $0.5$ ($e=0.14$). Therefore the complexity of the system depends on the basin size of the synchronous resonance when such solution is stable. For $e>0.18$, the entropy starts to follow the $2/2$ SOR. Hence in general the entropy acts in accordance to the basin size of the SOR with lowest order available.

\section{Moon \& Mercury}
\label{sec:moon_and_mercury}

We now proceed to the cases of the Moon and Mercury, meaning that we choose orbital eccentricities that correspond to those of these celestial bodies, namely $e_{Moon}=0.0549$ and $e_{Mercury}=0.2056$, and we set a constant parameter which is closer than the previous case to a real-life scenario but high enough so that numerical simulations can be carried out in a feasible amount of time, namely $K=10^{-4}$. In both cases, we adopted an equatorial oblateness of $\gamma=10^{-4}$. 

\begin{figure*}[htbp]
\centering
  \includegraphics[scale=0.9]{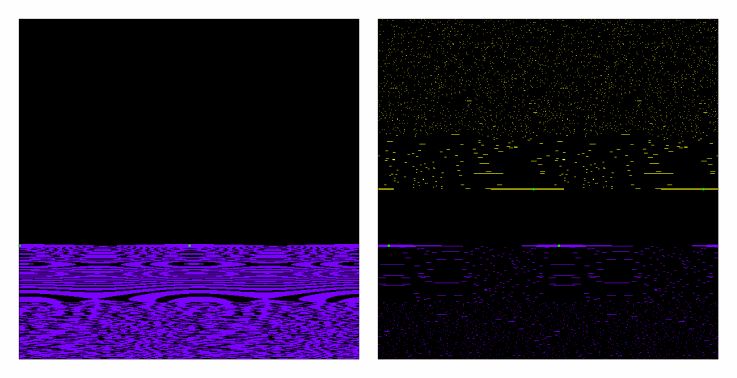}
\caption{Basins of attraction for (left) $e=e_{Moon}$ and (right) $e=e_{Mercury}$. Dissipation constant and equatorial oblateness are given by $K=10^{-4}$ and $\gamma=10^{-4}$, respectively. The color code corresponds to the following attractors: quasi-periodic (black), 1/1 (purple), and 3/2 (yellow). The location for the SORs is marked with little green dots.}
\label{fig:boa_moon_mercury}
\end{figure*}

In Fig.~\ref{fig:boa_moon_mercury}, we present the basins of attraction for the Moon (left) and Mercury (right). 
Here we set a fixed step size for the numerical integration scheme, and we calculated the basin of attraction on a square grid of $300\times300$ initial conditions. In both cases, there exists quasi-periodic attractors, whose basin is shown in black. As was done for Hyperion regarding higher-order resonances, we do not differentiate between distinct quasi-periodic attractors and consider them to represent the same final state. The basin for the synchronous resonance is painted purple, while the one for the 3/2 SOR is painted yellow.

The scenario for both the Moon and Mercury is quite different from Hyperion's (see Fig.~\ref{fig:boa_hyperion}). First, there is the presence of quasi-periodic attractors in the system. Second, the basin of attraction associated with the quasi-periodic attractor separates the other basins in the phase space. Hence, there is no basin boundary between the synchronous resonance and the 3/2 SOR for Mercury.

The compactness of the basins in Fig.~\ref{fig:boa_moon_mercury} can be associated to the lower value of $\gamma$, where the system behaves more like a pendulum, and the lower value of $K$, where the system does not differ much from a conservative system. In this situation, the manifolds that separate the phase space are packed closely together, leading to a dynamical scenario different from the one observed for Hyperion. 

\renewcommand{\arraystretch}{1.2}
\begin{table}[htbp]
\caption{Basin sizes and entropy for $\gamma=10^{-4}$, $K=10^{-4}$, and the eccentricities of the Moon and Mercury.}
\label{tab:basin_sizes_moon_mercury}  
\centering
\begin{tabular}{ccccc}
\hline\noalign{\smallskip}
 & \multicolumn{3}{c}{Basin size} & Entropy\\
\hline\noalign{\smallskip}
e & 1/1 & 3/2 & Quasi-periodic & \\
\noalign{\smallskip}\hline\noalign{\smallskip}
$e_{Moon}$ & 22.13\% & & 77.87\% & 0.529\\
$e_{Mercury}$ & 1.38\% & 1.17\% & 97.45\% & 0.136\\
\noalign{\smallskip}\hline
\end{tabular}
\end{table}

While the Moon contains only the synchronous resonance besides the quasi-periodic attractor, Mercury also contains the 3/2 SOR. The associated basin sizes and entropy are shown in Tab.~\ref{tab:basin_sizes_moon_mercury}. Even though both situations present an intricate basin structure, the quasi-periodic attractor dominates the phase space for $e=e_{Mercury}$, and hence the entropy is smaller in comparison to the Moon, where the synchronous resonance fills a large area of the phase space.

\begin{figure*}[htbp]
\centering
  \includegraphics[scale=0.75]{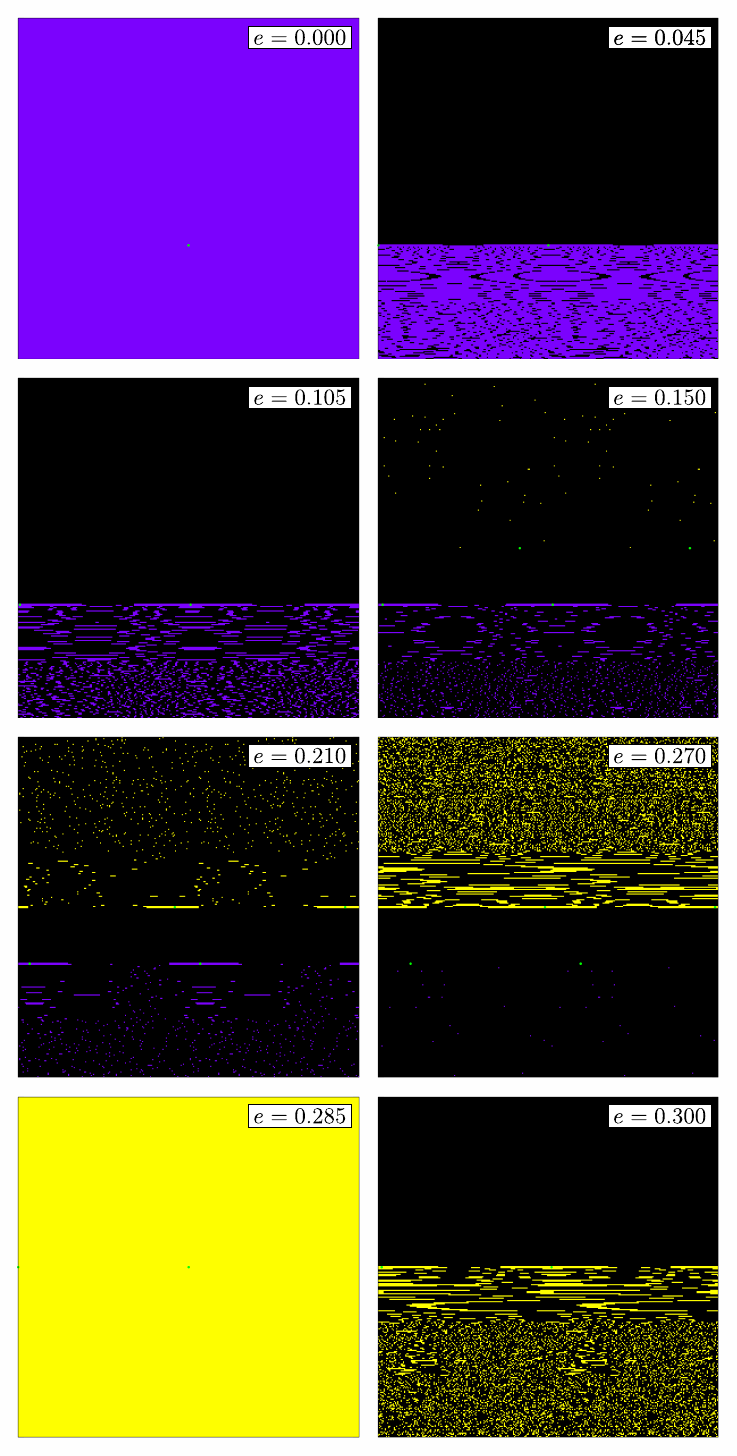}
\caption{Basins of attraction for $K=10^{-4}$, $\gamma=10^{-4}$, and different values of $e$. The color code corresponds to the following attractors: quasi-periodic (black), 1/1 (purple), and 3/2 (yellow). The location for the SORs is marked with little green dots.}
\label{fig:boa_low_gamma}
\end{figure*}

\begin{figure*}[htbp]
\centering
\includegraphics[scale=0.2]{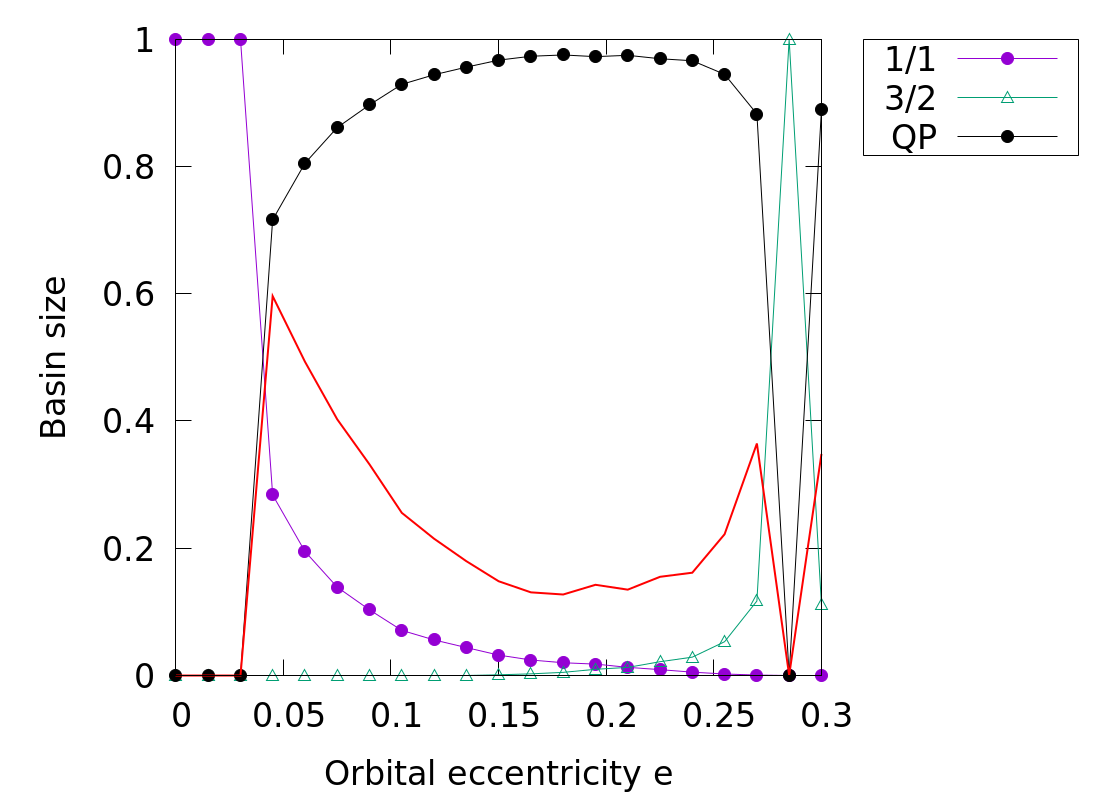}
\caption{Size of the basins of attraction for $K=10^{-4}$ and $\gamma=10^{-4}$, where QP stands for quasi-periodic attractors. The entropy is shown in red.}
\label{fig:BOA_size_low_gamma}
\end{figure*}

In Fig.~\ref{fig:boa_low_gamma}, we maintain $\gamma=10^{-4}$ and $K=10^{-4}$, and we depict the basins of attraction for different values of $e$. As before, the basin of the quasi-periodic solutions are shown in black, while the basin for the synchronous resonance is painted purple, and the one for the 3/2 SOR is painted yellow. The basin sizes and the corresponding entropy are shown in Fig.~\ref{fig:BOA_size_low_gamma}. For $e=0.0$, as expected, the synchronous resonance completely fills the phase space. For $e=0.045$, a quasi-periodic orbit appears and its basin dominates the phase space. As the orbital eccentricity is increased, the basin of the 3/2 SOR steadily grows, until finally covering all the phase space at $e=0.285$, where both the synchronous resonance and the quasi-periodic attractor disappear. After that, another quasi-periodic attractor appears along with its basin, which competes for space with the 3/2 SOR basin.

\section{Conclusions}
\label{sec:conclusions}

In this work, we have visually illustrated the basins of attraction for the dissipative spin-orbit problem, considering the parameters of Hyperion as a toy model, and the parameters for the Moon and Mercury. In all cases, the basins presented an intricate structure, being highly intertwined with one another. The visual depiction of the system's complexity was complemented with the measure of the Gibbs entropy in each case, which takes into consideration the number of attractors, and was shown to be correlated to the dominant attractor in the system.

For Hyperion, the probability of a trajectory ending up in a synchronous rotation not only varies considerably as a function of the orbital eccentricity, but also the 1/1 SOR itself bifurcates and changes stability, rendering such final state impossible for $e$ higher than a critical value. For the Moon and Mercury, where the equatorial oblateness is lower, quasi-periodic attractors appear and their basins of attraction act as barriers between the synchronous solution and the 3/2 SOR for values of $e$ where the latter attractor exists.

In the future, we intend to use a more realistic rheological model (see, for example, \citeauthor{Ragazzo2017} \citeyear{Ragazzo2017}), and analyse how changes in the equatorial oblateness, which are due to long term variations of spin and tidal forces, might affect the topology of the basins of attraction. It is worth noticing that the Gibbs entropy investigated here can be easily extended to higher-dimensional physical models, which is important since we do not have access to tools for visually describing the system in this situation. Furthermore, we also found it to be a good criterion for convergence in Monte-Carlo simulations (see Appendix).

\section*{Data availability}

The datasets generated and/or analysed during the current study are available from the corresponding author on reasonable request.
The software used to generate and analyse said datasets was written by the authors and it is available at \url{https://github.com/vitor-de-oliveira/spin-orbit}. 

\section*{Conflict of interest}

The authors declare that they have no known competing financial interests or personal relationships that could have appeared to influence the work reported in this paper.

\section*{Acknowledgements}
VMO would like to thank Clodoaldo Ragazzo for his comments and suggestions. VMO would also like to thank Project Gutenberg for making a book referenced in this article available online. This study was financed by the São Paulo Research Foundation (FAPESP, Brazil), under Grants No. 2021/11306-0 and 2022/12785-1.

\section*{Appendix: Monte Carlo simulations for lower dissipation values}

The dissipation constant sets the timescale for energy loss in the system, hence lower values of $K$ means that orbits on the phase space take longer to converge on average, and computational time can become prohibitive in this situation. However, it is often possible to obtain results using Monte-Carlo simulations, since these require less initial conditions in order to measure a given quantity. In fact, we did compare our computations on the grid to Monte-Carlo simulations, for which we obtained similar results. A Monte-Carlo simulation is faster but the uniform grid method allows us to visually depict the basins of attraction, which was the focus of our work.

\begin{figure*}[htbp]
\centering
\includegraphics[scale=0.87]{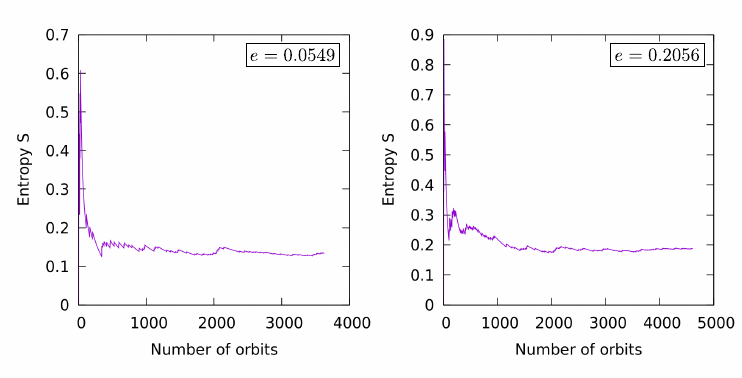}
\caption{Entropy as a function of the total number of randomly selected orbits for ${\gamma}=10^{-4}$, $K=10^{-6}$,  (left) $e=e_{Moon}$, and (right) $e=e_{Mercury}.$ The convergence of the measure $S$ was used as a criterion for the convergence of the Monte-Carlo simulations.}
\label{fig:entropy_progress}
\end{figure*}

In Tables~\ref{tab:basin_sizes_moon_K}~and~\ref{tab:basin_sizes_mercury_K}, we present the basin sizes and the entropy for values of $K$ equal and lower than the one adopted in Sec.~\ref{sec:moon_and_mercury}. In each simulation, we have selected up to $10^{4}$ random initial conditions, evolved them in time, and used the entropy as a convergence criterion for the method (see Fig.~\ref{fig:entropy_progress}). In the  regime of lower values of the dissipation constant, there are more stable solutions whose basins compete for space. However, in almost all scenarios, the basins of attraction of quasi-period attractors almost completely dominated the observed region of the phase space.

\renewcommand{\arraystretch}{1.2}
\begin{table}[htbp]
\caption{Basin sizes and entropy for $\gamma=10^{-4}$ and $e=e_{Moon}$.}
\label{tab:basin_sizes_moon_K}  
\centering
\begin{tabular}{cccccc}
\hline\noalign{\smallskip}
 & \multicolumn{4}{c}{Basin size} & Entropy\\
\hline\noalign{\smallskip}
K & 1/1 & 3/2 & Quasi-periodic & Other SORs & \\
\noalign{\smallskip}\hline\noalign{\smallskip}
$10^{-4}$ & 21.64\% &  & 78.36\% &  & 0.522 \\
$10^{-5}$ & 0.91\% & 0.41\% & 98.68\% &  & 0.079 \\
$10^{-6}$ & 0.85\% & 0.61\% & 97.91\% & 0.63\% & 0.124 \\
\noalign{\smallskip}\hline
\end{tabular}
\end{table}

\renewcommand{\arraystretch}{1.2}
\begin{table}[htbp]
\caption{Basin sizes and entropy for $\gamma=10^{-4}$ and $e=e_{Mercury}$.}
\label{tab:basin_sizes_mercury_K}  
\centering
\begin{tabular}{cccccc}
\hline\noalign{\smallskip}
 & \multicolumn{4}{c}{Basin size} & Entropy\\
\hline\noalign{\smallskip}
K & 1/1 & 3/2 & Quasi-periodic & Other SORs & \\
\noalign{\smallskip}\hline\noalign{\smallskip}
$10^{-4}$ & 1.21\% & 1.27\% & 97.52\% &  & 0.133 \\
$10^{-5}$ & 0.94\% & 0.43\% & 98.28\% & 0.35\% & 0.132 \\
$10^{-6}$ & 0.54\% & 0.65\% & 97.07\% & 1.74\% & 0.160 \\
\noalign{\smallskip}\hline
\end{tabular}
\end{table}

\bibliographystyle{myspbasic}      

\bibliography{manuscript_citation}   

\end{document}